\documentstyle[preprint,aps]{revtex}

\newcommand{\gsim}{\mbox{\raisebox{-1.0ex}{$\stackrel{\textstyle >}
{\textstyle \sim}$ }}}
\newcommand{\lsim}{\mbox{\raisebox{-1.0ex}{$\stackrel{\textstyle <}
{\textstyle \sim}$ }}}

\newcommand{\beq}{\begin{equation}}
\newcommand{\eeq}{\end{equation}}
\newcommand{\beqa}{\begin{eqnarray}}
\newcommand{\eeqa}{\end{eqnarray}}

\newcommand{\tot}{3/2}
\newcommand{\mpl}{M_{Pl}}

\newcommand{\lkk}{\left[}
\newcommand{\rkk}{\right]}

\tightenlines
\begin{document}
\thispagestyle{empty}
\makeatletter
\def\@preprint{
\noindent\hfill\hbox{YITP/U-93-29}\vskip 1pt
\noindent\hfill\hbox{LBL-34887}\vskip 10pt}
\makeatother

\title{Chaotic Inflation and Baryogenesis in Supergravity\thanks{This
work was supported in part by the Japanese Grant-in-Aid for 
Scientific Research Fund of Ministry of Education, Science, and 
Culture, Nos.\ 05740276 and 05218206, and in part by the
Director, Office of Energy 
Research, Office of High Energy and Nuclear Physics, Division of High 
Energy Physics of the U.S. Department of Energy under Contract 
DE-AC03-76SF00098.}}

\author{Hitoshi Murayama\cite{leave1}}
\address{Theoretical Physics Group, Lawrence Berkeley Laboratory\\
University of California, Berkeley, CA 94720, USA}
\author{Hiroshi~Suzuki\cite{leave2}}
\address{Instituto Nazionale di Fisica Nucleare, Sezione di Genova,
  16146 Genova, Italia}
\author{T.~Yanagida}
\address{Department of Physics, Tohoku University, Sendai, 980 Japan}
\author{Jun'ichi~Yokoyama}
\address{Uji Research Center, Yukawa Institute for Theoretical Physics\\
Kyoto University, Uji, 611 Japan}
\maketitle

\begin{abstract}
We propose a K\"ahler potential in supergravity which successfully
accommodates chaotic inflation. This model can have a large gravitino
mass without giving a large mass to  squarks and sleptons, and thus is
free  from both the gravitino problem and entropy crisis. 
In this  model  baryogenesis takes place naturally, identifying
the inflaton with a right-handed sneutrino with its mass 
$M \simeq 10^{13}$GeV, which is consistent  with the COBE data and the 
Mikheyev--Smirnov--Wolfenstein solution to the solar neutrino problem.
The model can also accommodate the matter content appropriate for the
mixed dark matter scenario.
\end{abstract}

\newpage
\renewcommand{\thepage}{\roman{page}}
\setcounter{page}{2}
\mbox{ }

\vskip 1in

\begin{center}
{\bf Disclaimer}
\end{center}

\vskip .2in

\begin{scriptsize}
\begin{quotation}
This document was prepared as an account of work sponsored by the United
States Government.  Neither the United States Government nor any agency
thereof, nor The Regents of the University of California, nor any of their
employees, makes any warranty, express or implied, or assumes any legal
liability or responsibility for the accuracy, completeness, or usefulness
of any information, apparatus, product, or process disclosed, or represents
that its use would not infringe privately owned rights.  Reference herein
to any specific commercial products process, or service by its trade name,
trademark, manufacturer, or otherwise, does not necessarily constitute or
imply its endorsement, recommendation, or favoring by the United States
Government or any agency thereof, or The Regents of the University of
California.  The views and opinions of authors expressed herein do not
necessarily state or reflect those of the United States Government or any
agency thereof of The Regents of the University of California and shall
not be used for advertising or product endorsement purposes.
\end{quotation}
\end{scriptsize}

\vskip 2in

\begin{center}
\begin{small}
{\it Lawrence Berkeley Laboratory is an equal opportunity employer.}
\end{small}
\end{center}

\newpage
\renewcommand{\thepage}{\arabic{page}}
\setcounter{page}{1}

Cosmic inflation is regarded as a necessary ingredient in the evolution
of the early universe not only to solve the horizon and the flatness
puzzles \cite{Guth} as well as the monopole \cite{Sato}, the
domain-structure \cite{Ksato} and the gravitino problems \cite{Weinberg}
but also to account for the origin of density fluctuations \cite{yuragi}
as observed by the Cosmic Background Explorer (COBE) satellite
\cite{cobe}. 
Among the various types of the inflationary models proposed so far, 
the chaotic inflation \cite{Linde} is the simplest 
mechanism to realize the accelerated expansion, which can 
accommodate inflation even in the presence of large quantum fluctuation
in the very early universe.

On the other hand, the idea of inflation generally depends on the
existence of the extremely high energy scale, typically of the order of
grand-unification scale or Planck scale. This introduces a huge
hierarchy between the weak scale and the high energy scale, and the
Higgs potential in the standard model becomes unstable against the
radiative corrections. The only solution to ensure the stability of the
hierarchy in the scalar potential against the radiative corrections is
the supersymmetry (SUSY) \cite{hierarchy}. Furthermore in chaotic inflation,
one has to deal with the scalar fields beyond $\mpl$, where supergravity
effects will be important.

Unfortunately, these two ideas, chaotic inflation and supergravity,
have not been 
naturally realized simultaneously so far. The main reason is
that ``minimal'' supergravity potential has an expotenential factor
which prevents any scalar fields from acquiring values larger than
$\mpl$ even in the chaotic epoch. 

Furthermore, the supergravity
model suffers from other difficulties and constraints as listed below.
When one adopts the SUSY breaking in the hidden sector, 
one usually encounters a light scalar field with mass of the order of
the gravitino mass $m_{\tot}$, and its value at the potential minimum is
$O(\mpl)$. It rolls down to the minimum only after $H \sim m_{\tot}$, and
produces huge entropy with a dilution factor of $O(10^{15})$
\cite{Coughlan}.  It was also recently pointed out that the phenomenon is
general in a wide class of superstring inspired models \cite{Nelson}.
Unlike the gravitino problem, inflation provides no help
to solve such an entropy crisis. 
Of course, we  should also  have a vanishing
cosmological constant at the potential minimum and be able to 
accommodate TeV scale squarks/sleptons masses.

In the present letter, we propose a K\"ahler potential as a model 
of chaotic inflation in
supergravity in which  all the problems cited above are resolved.
We will also discuss that the inflaton can be identified with the
right-handed sneutrino, which was proposed in our previous publication
\cite{MSYY}. This idenfication allows us to generate baryon asymmetry
in the reheating process, and can naturally accommodate the mixed dark
matter scenario. 

Before proceeding to our potential, in order to confirm the
difficulty in implementing the chaotic inflation into supergravity as
mentioned above, let us recall 
the Lagrangian for scalar multiplets $\varphi^i$ in generic supergravity
models,
\beq
{\cal L_\varphi}=G_i^j\partial_\mu \varphi^i \partial^\mu \varphi^*_j
-e^G\lkk G_i(G^i_j)^{-1}G^j-3\rkk, \label{scalarl}
\eeq
where $G_i=\partial G/\partial \varphi^i$, 
$G^j=\partial G/\partial \varphi^*_j$, and we have taken
a unit with $\mpl/\sqrt{8\pi}=1$.
In the ``minimal'' supergravity, to ensure the canonical 
kinetic term for $\varphi$,
one takes the K\"ahler potential as 
$G[\varphi, \varphi^*]=\varphi^i\varphi^*_i + \ln|W[\varphi]|^2$,
where $W[\varphi]$ is the superpotential.
Then the scalar potential has an exponential factor 
$e^{\varphi_i^* \varphi^i}$ which makes it very
difficult to incorporate chaotic inflation.  In fact, as
suggested by Goncharov and Linde, one must introduce either a rather
specific class of nonpolynomial superpotential \cite{glpl} or some
fields with nonminimal kinetic term \cite{glcqg}. In the former model,
one has to
arrange the parameters so that the potential is bounded from below while
it does not grow rapidly to enable chaotic inflation. It also has the
same entropy crisis as in the minimal supergravity.
We have also found
that the latter model \cite{glcqg} suffers from a runaway behavior of
the inflaton, since the imaginary part of the inflaton has negative mass
squared when the real part is of order of the Planck mass. Then the
imaginary part grows without bound,  and the scalar potential for the
matter fields damps exponentially leading to a vanishing potential.

Now we propose the following K\"ahler potential free from such 
difficulties, 
\beqa
G &=& \frac{3}{8} \ln \eta + \eta^2 + \ln |W(\varphi)|^2,
\label{gpot}\\
\eta &=& z + z^* + \varphi_i^* \varphi^i,
\eeqa
where $\varphi^i$ denotes chiral fields in the model.
Then the scalar Lagrangian is given by \cite{Gelmini}
\beq
{\cal L} = \frac{16 \eta^2 - 3}{32 \eta^2}
	\left[ (\partial_\mu \eta)^2 + (I_\mu)^2 \right]
	+ \frac{16 \eta^2 + 3}{8\eta} |\partial_\mu \varphi^i|^2
	- V,
\eeq
with the potential 
\beqa
V &=& \eta^{3/8} e^{\eta^2} \left\{
	\frac{8\eta}{16\eta^2 + 3} |W_i|^2
	+ \frac{(16 \eta^2-9)^2}{8(16 \eta^2 -3)} |W|^2 \right\}
		\nonumber\\
& &	+ D\mbox{-terms}
\label{potential}
\eeqa
where $W_i = \partial W/\partial \varphi^i$,
and $I_\mu$ is a $U(1)$ current defined by
$I_\mu = i \partial_\mu(z - z^*) + 
	i (\varphi^i \partial_\mu \varphi_i^* -
		\varphi_i^* \partial_\mu \varphi^i) .$

Figure 1 depicts the shape of the potential along the $\eta$-axis.
The kinetic term of $\eta$ has the correct signature for $|\eta|>\sqrt{3}/4$.
The minimum of the potential $V$ for $\eta >\sqrt{3}/4$ is given by
$\eta=3/4$ and
$W_i= 0$,
where the  cosmological constant vanishes as desired. Note that the
potential Eq.~(\ref{potential}) has exactly the same form with the
global SUSY case once $\eta$ has settled to its minimum.
Therefore, this K\"ahler potential realizes chaotic inflation, once one
has a sensible model within global SUSY. On the other hand, 
this scalar potential also has flat directions, which might be
cosmologically harmful. We return to this point later.

The SUSY is broken down by adding a constant term in the superpotential.
The gravitino mass is given by
\beq
m^2_{\tot} = e^G = \eta^{3/8} e^{\eta^2} |W|^2
\eeq
In order to give the gravitino a mass of $m_{3/2}$
at the potential minimum,
we simply add a constant of
$O(m_{\tot})$ in the superpotential. 
Note that the constant does not induce the soft SUSY breaking terms to
$\varphi^i$. Therefore, one can choose the constant freely 
without giving large masses to the squarks and sleptons. On the other
hand, one 
can give gaugino mass $m_\lambda$, assuming the form of the kinetic
function $f_{ab} = \delta_{ab} h(z)$,
\beq
m_\lambda = m_{\tot} 
	\frac{\eta (16 \eta^2 + 3)}{2(16\eta^2 -3)}
	\frac{\partial h/\partial z}{\Re e(h(z))}
	= \frac{3}{4} m_{\tot} 
		\left. \frac{\partial h/\partial z}{\Re e(h(z))}
		\right|_{z = \frac{3}{8}}
\eeq
at the potential minimum $\eta = 3/4$. The denominator $\Re e(h(3/8))$
should be $O(1)$ to give correct gauge coupling constants, 
while one can  choose its derivative in the numerator to set gaugino
mass below 10 TeV so that radiative corrections do not induce too large
masses to squarks and sleptons \cite{loop}.

Now let us turn to cosmological consideration. It is evident that 
inflation is possible only in the domain with $\eta>\sqrt{3}/4$ initially, 
which we shall assume hereafter.
Recall the criterion by Linde \cite{Linde} that the scalar fields may
have any initial configurations at the Planck time as far as the energy
density is less than $O(1)$ in the Planck unit.
We expect that the initial amplitude of the gauge non-singlet fields at
the Planckian epoch are of order unity due to the existence of the
$D$-term potential \cite{finetuning}.
Also, $\eta$ has $O(1)$ fluctuations from its minimum at that epoch.
A gauge-singlet field $\phi$ can have, on the other hand, a value larger than
$O(1)$ if its potential $|W_i|^2$ is sufficiently flat, and it can
drive chaotic inflation. The simplest possibility is 
\beq
	W = \frac{1}{2} M \phi^2
\eeq
where the mass $M$ should be $M \simeq 10^{13}$~GeV to reproduce COBE
data \cite{Salopek}. Then it can have a large value $\phi(0) \simeq
1/\sqrt{M}$ even though other fields have $O(1)$ value, limited by the
last term $|W|^2$ in the potential. Then $\eta$ quickly settles to its
minimum to make $|W|^2$ term practically vanishing, and the potential
for the 
$\phi$ is nothing but the global SUSY potential, $M^2
|\phi|^2$. The other gauge non-singlet fields may drop either to
the origin, or one of the flat directions with $O(1)$ amptlidues at most.

The dynamics of the universe after its energy  is dominated by 
that of $\phi$ is the same as in the ordinary chaotic inflation scenario.
During the inflation, $\phi(t)$ and the scale factor $a(t)$
are given by 
\begin{eqnarray}
	\phi (t)
	&=& \phi(0)  - \frac{M}{\sqrt{3}}t,\\
a(t) &=& a(0) \exp \left[  \frac{M}{\sqrt{3}} 
		 \phi(0)t
	\left( 1 - \frac{Mt}
		{2\sqrt{3}\phi(0)} \right) 
	 \right],
\end{eqnarray}
respectively. At the time $t_e$, 
when $\phi(t_e) \simeq 1$, the exponential expansion ends. Assuming
the initial condition $\phi (0) \simeq 1/\sqrt{M}$, the duration of
the inflation is $t_e \simeq 1/M^{3/2}$, and the $e$-folding is $1/M
\simeq 10^{5}$. The necessary initial condition to solve the horizon and
flatness problems is actually only  $\phi(0) \simeq 15$. 

Now we discuss the possible entropy production in our model. The
potentially most dangerous field is $\eta$. This field is a
superpartner of the goldstino, and therefore one may worry about
the huge entropy production \cite{Coughlan,Nelson}. Though the true
minimum is at $\eta = 3/4$, the minimum for $\eta$ depends on the values
of $|W|^2$ and $|W_i|^2$ during the inflation. The Fig.~2 shows the
minimum of $\eta$ as a function of $|W_i|^2/|W|^2$. The minimum of
$\eta$ was at the true minimum in the beginning of the inflation, but it
becomes gradually shifted to $\eta_e \simeq 0.70$ at the end of the
inflation. It stays there until the expansion rate becomes as small as
its mass. If we assume its mass around $O(1)$~TeV, it may produce
enormous amount of entropy of $O(10^{15})$ dilution factor 
as mentioned in the beginning.
Fortunately, our model allows us to have much larger
gravitino mass $m_{3/2}$, and simultaneously also for $\eta$, without
giving large mass to squarks/sleptons. 
Then the decay of the coherent
oscillation of $\eta$ produces a dilution factor of
$(\Gamma_\phi/\Gamma_\eta)^{1/2} (\eta_e-3/4)^2/3$, which is $O(1)$
unless $\Gamma_\eta \simeq m_{\tot}^3 \ll 10^{-6} \Gamma_\phi$.
Therefore our model can avoid the entropy crisis
by taking an appropriate
particle-physics model of the inflaton $\phi$. We will return to this
point later.
Moreover 
the gravitinos heavier than $10^{8}$~GeV
decay well before the electroweak phase transition, and are
completely harmless; there is no gravitino problem at all \cite{ENQ}.

Another possible source of entropy is the flat direction in the SUSY
potential whose mass is only of the order of TeV, and the scalar fields
may grow along that direction by the quantum fluctuation during the de
Sitter expansion of the universe \cite{BD}.
Though the duration of the inflation, $t_e \simeq M^{-3/2}$,
 is long enough to solve horizon
and flatness problems, it is too short for the flat direction to
grow beyond the Planck scale. The growth of the flat direction field,
$\chi$, by the quantum fluctuation is
known to be $\langle \chi^2 \rangle = (H^3/4\pi^2) t_e$ in the de Sitter
space with constant expansion rate $H$. 
In our model, the exponential expansion begins when the
potential is dominated by $M^2 |\phi|^2$ term with $\phi
\lsim 1/\sqrt{M}$, and the maximum expansion rate is $H \lsim
\sqrt{M}$. Therefore, the amplitude of quantum
fluctuations is much smaller than unity;
$\sqrt{\langle \chi^2 \rangle} \lsim 1/(4\sqrt{3}\pi) \ll 1$. 
Thus the only fate of the flat direction is
to decay with the decay rate $\Gamma \simeq \alpha m (m/\chi)^{1/2}$
\cite{DK} giving a dilution factor of $O(10^{2})$ at maximum even if 
$\chi = O(1)$ classically at the outset \cite{2nd}.

Finally, we will comment below that this model can successfully
incorporate baryogenesis if one identifies $\phi$ with the right-handed
sneutrino $\tilde{N}$ \cite{MSYY}.
The right-handed neutrino is naturally introduced from particle-physics
point of view to account for a small
neutrino mass required for the 
Mikheyev--Smirnov--Wolfenstein (MSW) solution
\cite{MSW} to the solar neutrino puzzle \cite{solar}
by the seesaw mechanism \cite{seesaw}.
Since it is a gauge-singlet field with small Yukawa coupling at least
for the first generation, 
the sneutrino can naturally drive chaotic inflation.
The mass $M \simeq 10^{13}$GeV which was necessary to reproduce COBE
data nicely fits into the window for the MSW solution (see, {\it e.g.},\/
\cite{window}).
It can also account for baryogenesis quite naturally by first 
producing lepton asymmetry \cite{FY}
in the inflaton decay and later converting it partially to baryon asymmetry 
through the anomalous electroweak process 
\cite{FY,KRS,Arnold}. 
The entropy crisis can be also avoided when its Yukawa coupling is
$\lsim O(10^{-5})$ ($\Gamma_N \lsim 10^{2}$~GeV) 
and $m_{3/2} \gsim 10^{11}$~GeV ($\Gamma_\eta \gsim 10^{-3}$~GeV).

The resulting
baryon-to-entropy ratio $Y_B$ is \cite{MSYY}
\begin{equation}
Y_B = 0.35 \epsilon \frac{3T_{RH}}{4M}
	\exp \left[ - 0.012 \frac{G_F^2 m_{\nu_\tau}^2}{\sin^4 \beta}
		\mpl T_{RH} \right],
\end{equation}
where $T_{RH} = O(10^9$--$10^{10})$~GeV is the reheating temperature in
this model, $\epsilon$ the CP-violating decay asymmetry into leptons
and anti-leptons, $G_F$ the Fermi constant, $m_{\nu_\tau}$ mass of the
$\tau$-neutrino, and $\beta$ is the vacuum angle $\tan \beta=\langle H_u
\rangle /\langle H_d \rangle$. The exponential factor comes from the
annihilation of the lepton numbers in the thermal bath. 
To account for the desired $Y_B$ in the nucleosynthesis, one can
take small $\epsilon \simeq 10^{-8}$ with $m_{\nu_\tau} \lsim 2~$eV, but
it requires small CP-violating phase in the Yukawa coupling matrix. More
interesting case is that with $O(1)$ CP-violating phase;
{\it e.g.}\/, $\epsilon \simeq 10^{-3}$, $T_{RH} \sim 10^9$~GeV, and
$m_{\nu_\tau} \simeq 10$~eV \cite{blowup} gives desired $Y_B$ in
accordance with the nucleosynthesis. 
This means that our model
does not only accommodate the material ingredients for the mixed dark
matter model \cite{Mdavis} but in fact it {\it prefers}\/ that, with the 
cold dark matter component consisting of the lightest superparticle.
We also mention that the above mechanism is the only intrinsic 
source of baryogenesis in our model since
the flat direction cannot give rise to $B-L$ asymmetry by the
Affleck--Dine mechanism \cite{AD} in the presence of the large amplitude
of $\tilde{N}$ \cite{MSYY}.

JY acknowledges support by the Research Aid of Inoue Foundation for
Science.  This work was supported in part by 
the Japanese Grant-in-Aid for 
Scientific Research Fund of Ministry of Education, Science, and 
Culture, Nos.\ 05740276 and 05218206.
HM acknowledges support by the Director, Office of Energy 
Research, Office of High Energy and Nuclear Physics, Division of High 
Energy Physics of the U.S. Department of Energy under Contract 
DE-AC03-76SF00098.

\newpage
\section*{Figure Captions}
\renewcommand{\labelenumi}{Fig.~\arabic{enumi}}
\begin{enumerate}
\item The potential Eq.~(\ref{potential}) as a function of $\eta$, when
$W_i = 0$ for all the other fields. 
\item The value of $\eta$ at the minimum of the potential
Eq.~(\ref{potential}), as a function of $|W_i|^2/|W|^2$.
\end{enumerate}


\begin{references}
\bibitem{finetuning}
 Though it is possible in principle that the gauge non-singlet fields
 have larger value than unity along the $D$-term flat directions, it is
 unlikely in the chaotic configuration of the universe at the
 Planck-epoch since it requires a fine-tuning between various scalar fields.
\bibitem{2nd}
 If the scalar fields could grow beyond the Planck scale during the
 inflation, it might have caused another inflation, and only tiny
 density fluctuation of $O(10^{-15})$ was possible. This discussion
 shows our model is free from this danger as well.
\bibitem{blowup}
 Under the condition that $h_{33}$ does not
 blow up below the Planck scale, $m_{\nu_\tau}$ can be 
 as large as 10 eV with $\sin\beta=1$ and $M=10^{13}$ GeV.
\end{references}
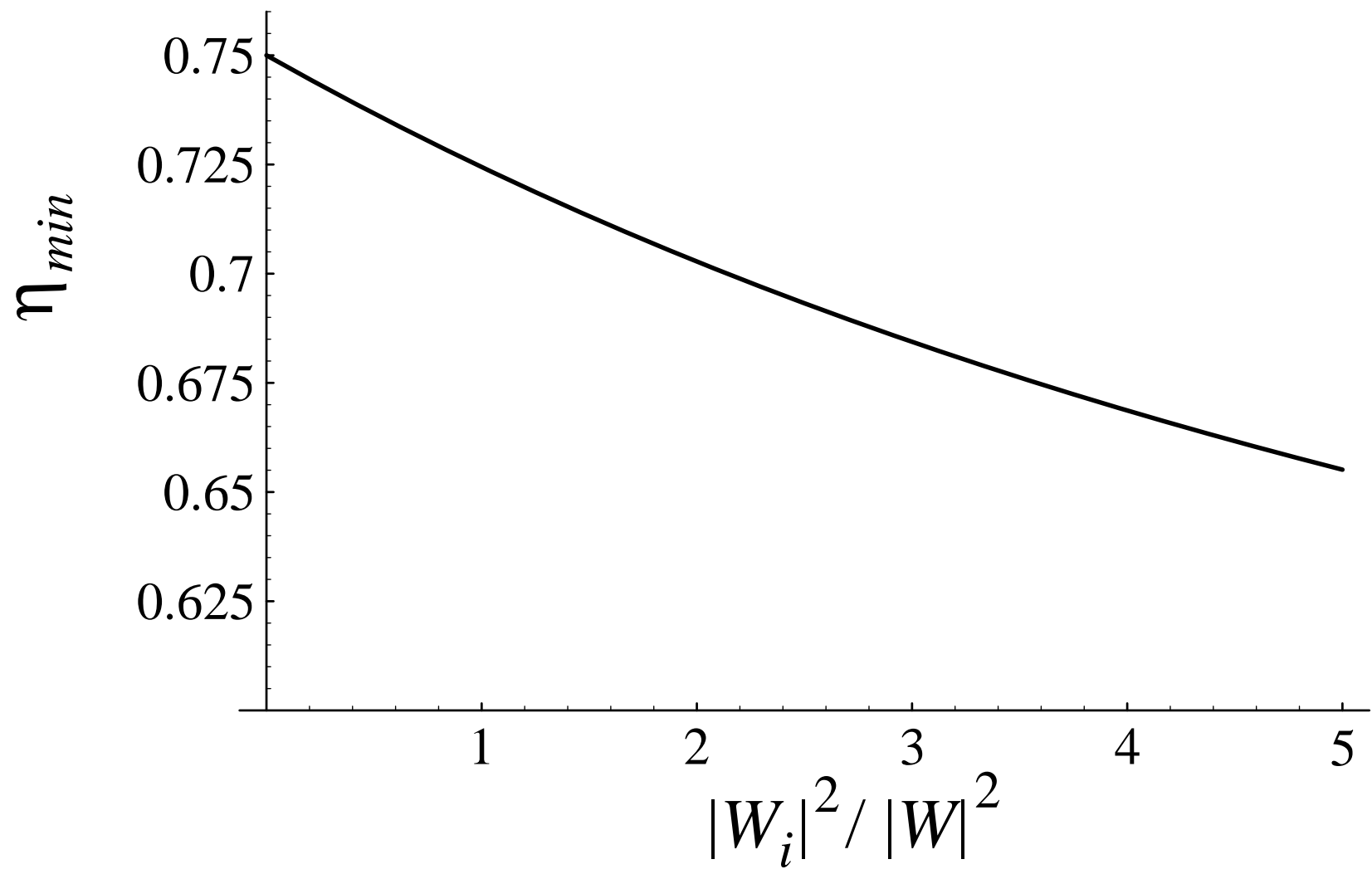
\end{document}